\PassOptionsToPackage{dvipsnames}{xcolor}
\documentclass[10pt,nonacm,sigconf,screen]{acmart}

\usepackage[all]{nowidow}
\usepackage{xcolor}
\usepackage{subcaption}
\usepackage{hyperref}
\usepackage{acronym}
\usepackage{balance}

\usepackage[frozencache]{minted}
\usepackage[capitalise,noabbrev]{cleveref}
\crefformat{section}{\S#2#1#3} 
\crefformat{subsection}{\S#2#1#3}
\crefformat{subsubsection}{\S#2#1#3}
\crefrangeformat{section}{\S#3#1#4 to~\S#5#2#6}
\crefrangeformat{subsection}{\S#3#1#4 to~\S#5#2#6}
\crefrangeformat{subsubsection}{\S#3#1#4 to~\S#5#2#6}

\newcommand{\sysname}{\textsf{Enoki}}

\begin{document}

\author{Tobias Pfandzelter}
\affiliation{%
    \institution{TU Berlin \& ECDF}
    \city{Berlin}
    \country{Germany}}
\email{tp@mcc.tu-berlin.de}

\author{David Bermbach}
\affiliation{%
    \institution{TU Berlin \& ECDF}
    \city{Berlin}
    \country{Germany}}
\email{db@mcc.tu-berlin.de}

\title{\sysname{}: Stateful Distributed FaaS from Edge to Cloud}

\begin{abstract}
    Function-as-a-Service (FaaS) is a promising paradigm for applications distributed across the edge-cloud continuum.
    FaaS functions are stateless by nature, leading to high elasticity and transparent invocation.
    Supporting stateful applications, however, requires integrating data storage in FaaS, which is not trivial in an edge-cloud environment.

    We propose \sysname{}, an architecture for stateful FaaS computing replicated across the edge-cloud continuum.
    \sysname{} integrates a replicated key-value store with single-node FaaS systems at edge and cloud nodes in order to provide low-latency local data access for functions without breaking the abstraction of the FaaS programming model.
    We evaluate \sysname{} with microbenchmarks on an open-source prototype and demonstrate building a stateful FaaS application with multiple functions distributed over edge and cloud.
\end{abstract}

\maketitle

\section{Introduction}
\label{sec:introduction}

Elastic resource allocation, scale-to-zero, high composability, and the event-driven programming model make Function-as-a-Service (FaaS) a promising paradigm for applications on the edge-cloud continuum~\cite{paper_pfandzelter2020_tinyfaas,paper_pfandzelter2019_functions_vs_streams,jonas2019cloud,raith2023serverless}.
In FaaS, applications are composed of small, stateless functions that are executed in response to events.
Their infrastructure is completely abstracted by an underlying FaaS platform that manages dynamic resource allocation for function instances.

A key inhibitor to broad FaaS adoption is that many applications require some form of state management and can thus not be implemented in FaaS without relying on external services.
In the cloud, integrating a database service into a FaaS application is straightforward~\cite{paper_pfandzelter2022_streamingfunctions}.
At the edge, however, database access from a function should be local rather than incurring additional communication delay to the cloud.
For a distributed FaaS application, i.e., a function instantiated at multiple edge locations, this requires state synchronization and replication.
Crucially, to not break the abstraction of the FaaS paradigm, this should be transparent to applications.

To enable local data access for functions deployed in the edge-cloud continuum, we propose integrating an edge FaaS platform with a data management middleware.
To this end, we make the following contributions in this paper:

\begin{itemize}
    \item We lay out the architecture of \sysname{}, a FaaS platform for the edge-cloud continuum that integrates a distributed data replication service (\cref{sec:design}).
    \item We present an open-source proof-of-concept prototype for \sysname{} (\cref{sec:prototype}).
    \item We use our \sysname{} prototype to evaluate our approach in micro-benchmarks (\cref{sec:evaluation}).
    \item We then demonstrate running a multi-function edge-cloud IoT application on our \sysname{} prototype (\cref{sec:befaas}).
\end{itemize}

\section{\sysname{} Architecture}
\label{sec:design}

\begin{figure}
    \centering
    \includegraphics[width=\linewidth]{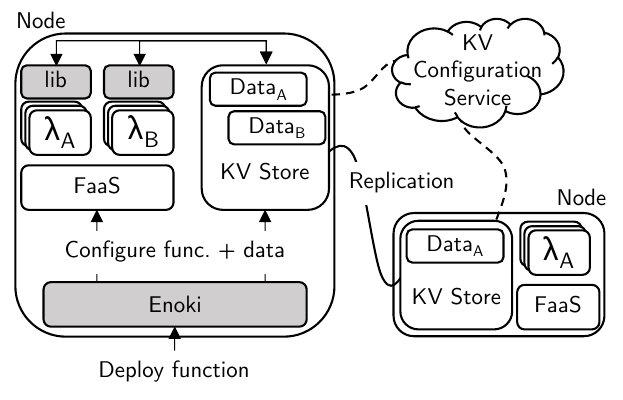}
    \caption{\sysname{} combines a FaaS platform with a replicated key-value (KV) store at each node. This allows functions to access data locally at edge nodes without sacrificing replication of functions across geo-distributed edge nodes.}
    \label{fig:architecture}
\end{figure}

\sysname{} is a distributed edge-to-cloud FaaS platform that integrates a replicated key-value store, as shown in \cref{fig:architecture}.
\sysname{} allows functions to access data \emph{locally} using a simple key-value interface while ensuring (i)~data consistency between parallel functions instances on the same host and (ii)~data replication for function instances on multiple hosts.

In \sysname{}, each host runs a lightweight single-node FaaS platform and exposes a function deployment API.
When a client instructs the deployment of a function to that host, \sysname{} creates the necessary function instances and checks if a key-value data container is available for that function.
If such a data container exists, i.e., the function was already instantiated on another \sysname{} host, it is replicated to the local host, where data then becomes available for local access from functions.
If no data container exists, it is created.

\sysname{} relies on the abstractions provided by the \emph{FReD} data replication service~\cite{pfandzelter2023fred}:
In FReD, applications control the replication of key-value data containers, called \emph{keygroups}, across locations in the edge-cloud continuum.
FReD uses a central \emph{naming service} that stores global configuration information (control flow), but is not involved in data flow.
Through a client-side library, FReD also ensures client-centric consistency in the event of concurrent data updates at different nodes, but data may be stale.
Applications can also implement their own logic to resolve conflicts using, e.g., convergent replicated data types (CRDT)~\cite{shapiro2011conflict}.
A key benefit of FReD is its integration of heterogeneous data store backends, e.g., in-memory storage at constrained edge nodes and database services such as \emph{DynamoDB} in the cloud.

\begin{listing}
    \begin{minted}[python3,baselinestretch=1.1]{python}
    import kv
    def call(i: str) -> str:
        curr = kv.get(key="current")
        curr = curr + "Hello World!\n"
        kv.set(key="current", val=curr)
        return curr
    \end{minted}
    \caption{Example \sysname{} function in Python: after importing the \mintinline{python}{kv} library, the function can read and write data locally using a simple, familiar CRUD interface.}
    \label{listing:example}
\end{listing}

We show an example of using the key-value interface for a stateful Python function in \cref{listing:example}.
The \mintinline{python}{import kv} directive initializes the connection to the local key-value store instance and imports the necessary functions for reading and updating data.
Note that in most FaaS platforms the global imports remain active with the function instance and warm starts do thus not incur a performance penalty for setting up the database connection.
Within the scope of \mintinline{python}{call(i: str)}, which is the entrypoint called for every function invocation, necessary state can be \emph{read} with the \mintinline{python}{kv.get(key: str)} method and \emph{written} with the \mintinline{python}{kv.set(key: str, val: str)} method.
While this example uses Python, a similar interface may be provided for other programming languages.
In addition to the single-key CRUD interface, \emph{scan} operations can also be supported for more complex data manipulation.

\section{Prototype}
\label{sec:prototype}

To aid in the evaluation of our proposed design, we develop a proof-of-concept prototype of \sysname{}, which we make available as open-source software.\footnote{\url{https://github.com/OpenFogStack/enoki}}
Our prototype is based on FReD as a key-value store and the \emph{tinyFaaS} lightweight edge FaaS platform.
We extend the default Docker backend in tinyFaaS with custom functionality that creates a FReD keygroup on the local FReD instance for each function or replicates an existing keygroup.
\sysname{} injects library code to access that keygroup into function handlers in order to maintain the high level of abstraction the FaaS paradigm offers.
After starting the central \texttt{etcd} naming service, e.g., on a central cloud node, every edge node starts a local tinyFaaS instance with its own local FReD node, which in turn connects to the central naming service.
Our prototype supports functions in Python using a Python 3.11 runtime.
Functions can be synchronously or asynchronously invoked over, e.g., HTTP.

\section{Single Function Evaluation}
\label{sec:evaluation}

Using our prototype, we demonstrate \sysname{} using a single FaaS function (\cref{sec:evaluation:single}), evaluate its impact on throughput in bandwidth-constrained environments (\cref{sec:evaluation:scaling}), and assess the benefits and drawbacks of data replication (\cref{sec:evaluation:replication}).

\subsection{Single Stateful Function}
\label{sec:evaluation:single}

To demonstrate the basic functionality of stateful functions on \sysname{}, we deploy a FaaS function that calculates a moving average of incoming data on top of edge-cloud infrastructure.

\subsubsection*{Setup}

\begin{figure}
    \centering
    \includegraphics[width=\linewidth]{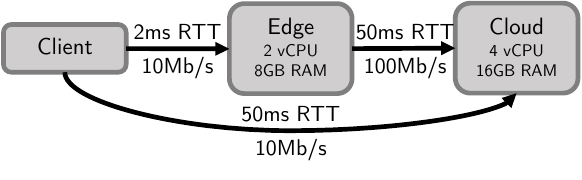}
    \caption{Our experiment setup emulates an edge-to-cloud environment with a client, a local edge node, and a more remote cloud node with larger compute capacity.}
    \label{fig:experiment-setup}
\end{figure}

Our simplified edge-cloud infrastructure setup shown in \cref{fig:experiment-setup} encompasses a client, an edge host, and a cloud server.
We deploy each machine as a Google Cloud Compute instance in the \texttt{europe-west9} (Paris, France) region, with the client and edge host of instance type \texttt{n2-standard-2} (2 vCPUs, 8GB RAM), and the cloud host of type \texttt{n2-standard-4} (4 vCPUs, 16GB RAM).
To emulate the characteristics of an edge-cloud network, we inject network delays and bandwidth restrictions between machines using the \texttt{tc-netem} network emulation tooling in Linux~\cite{brown2006traffic}.

In our experiments, we run \sysname{} on the edge node and then deploy our function on it.
By invoking this function from our client over HTTP, we compare request-response times between the database deployed in the cloud, as would traditionally be necessary, and at the edge with \sysname{}.
Note that we also use FReD on the cloud instance compared to a more cloud-native data store, e.g., DynamoDB, in order to maintain a fair comparison.

Our function takes an input, e.g., a sensor measurement by an IoT sensor, stores it in the key-value store, and then reads the ten most recent values from the store to compute a moving average.
In order to complete this, it also reads and updates a pointer to the most recent value.
Our client calls this function ten times per second (open workload~\cite{schroeder2006open}) for an experiment duration of five minutes.
We repeat each experiment three times to validate reproducibility.

\subsubsection*{Results}

\begin{figure}
    \centering
    \includegraphics[width=\linewidth]{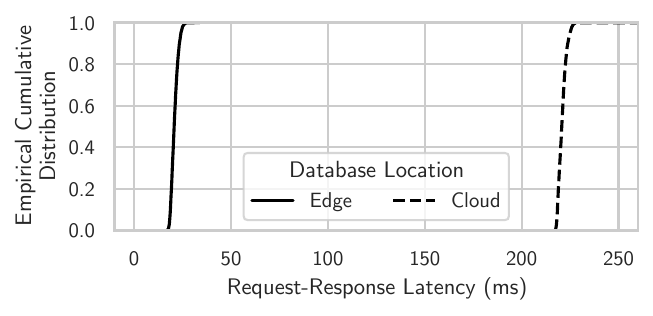}
    \caption{Request-response latency for the moving average function is increased by 200ms when reading and writing data from the remote cloud data store compared to local data access with \sysname{}.}
    \label{fig:single}
\end{figure}

The experiment results in \cref{fig:single} clearly show how reading and writing data to the cloud database adds a 200ms delay to the function computation.
As the data sizes read and written by our function are small (at most ten integer values read), this delay is likely caused only by the additional network delay, not by bandwidth constraints between edge and cloud.
With four requests for each function invocation (reading the pointer, scanning, writing the new value, and writing the new pointer) with an added 50ms each, the full function execution is thus extended by 200ms.

While not every connection between edge and cloud in the real world incurs a 50ms latency penalty, this experiment shows the direct impact of network distance on FaaS function performance for applications that depend on state storage.

\subsection{Read/Write Throughput}
\label{sec:evaluation:scaling}

Beyond network latency, bandwidth constraints between client, edge, and cloud can also impact the performance of distributed applications.
To assess the impact of these constraints, we measure the throughput of functions reading and writing data of varying size to the data store in \sysname{}.

\subsubsection*{Setup}

In this experiment we use the same infrastructure configuration as in \cref{sec:evaluation:single}.
We now deploy two functions to measure the impact of bandwidth constraints:
First, a read function reads data of a specified length from the key-value store (but does not return it to the client).
Second, a write function writes data of a specified length to the key-value store.
We again compare deployments of the key-value store on the edge and in the cloud, with the function itself always running on the edge host.
Our client uses a closed workload where 100 threads send requests to the edge function for two minutes.
We then measure system throughput based on data item size and completed tasks per second, with data sizes varied between one byte and one megabyte.
We repeat each experiments three times.

\subsubsection*{Results}

\begin{figure}
    \centering
    \begin{subfigure}{1\linewidth}
        \centering
        \includegraphics[width=1\linewidth]{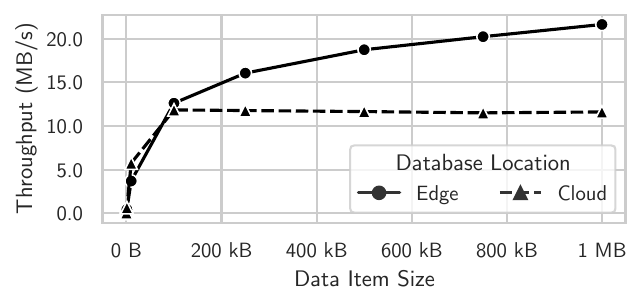}
        \caption{Read performance is limited by the 100Mb/s (12.5MB/s) edge-cloud bandwidth when locating the data store for our function in the cloud, while it increases with increased data item size for the edge location.
        }
        \label{fig:scaling:read}
    \end{subfigure}%
    \vfill
    \begin{subfigure}{1\linewidth}
        \centering
        \includegraphics[width=1\linewidth]{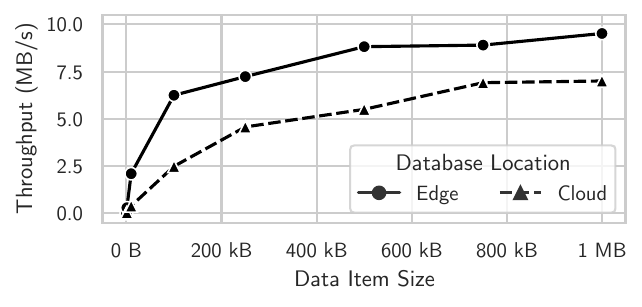}
        \caption{While throughput for both edge and cloud data store location converge with growing data size, performance for local data updates remains better than for remote data writing.}
        \label{fig:scaling:write}
    \end{subfigure}%
    \caption{Throughput performance of reading (\cref{fig:scaling:read}) and writing (\cref{fig:scaling:write}) data items of varying size is worse when accessing data in the cloud from an edge function than accessing data locally at the edge node.}
    \label{fig:scaling}
\end{figure}

\begin{figure*}
    \centering
    \begin{subfigure}{0.33\linewidth}
        \centering
        \includegraphics[height=2cm]{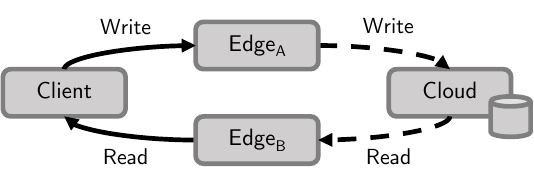}
        \caption{Cloud}
        \label{fig:replsetup:cloud}
    \end{subfigure}%
    \hfill
    \begin{subfigure}{0.33\linewidth}
        \centering
        \includegraphics[height=2cm]{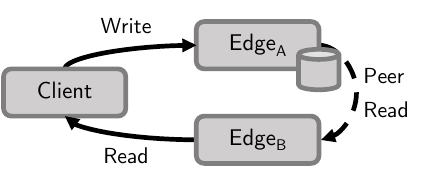}
        \caption{Edge (No Replication)}
        \label{fig:replsetup:p2p}
    \end{subfigure}%
    \hfill
    \begin{subfigure}{0.33\linewidth}
        \centering
        \includegraphics[height=2cm]{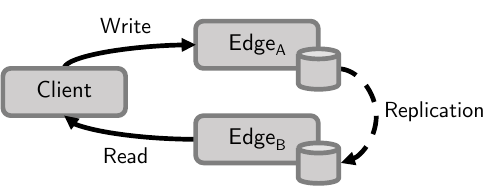}
        \caption{Edge (Replication)}
        \label{fig:replsetup:edge}
    \end{subfigure}%
    \caption{We compare cloud-based data access (\cref{fig:replsetup:cloud}), read access to a peer edge (\cref{fig:replsetup:p2p}), and edge-based local data access with replication (\cref{fig:replsetup:edge}) for read latency, write latency, and data staleness.}
    \label{fig:replsetup}
\end{figure*}

\begin{figure*}
    \centering
    \begin{subfigure}{0.33\linewidth}
        \centering
        \includegraphics[width=1\linewidth]{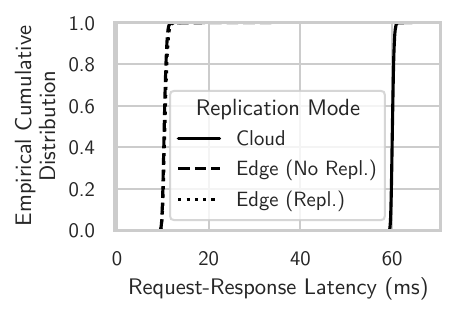}
        \caption{Write Operation}
        \label{fig:replication:write}
    \end{subfigure}%
    \hfill
    \begin{subfigure}{0.33\linewidth}
        \centering
        \includegraphics[width=1\linewidth]{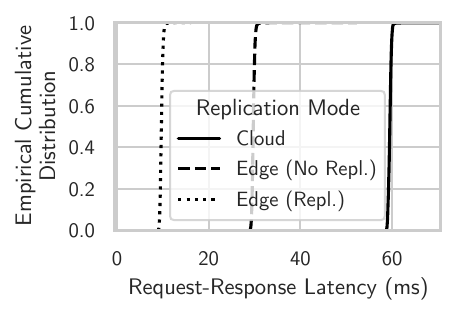}
        \caption{Read Operation}
        \label{fig:replication:read}
    \end{subfigure}%
    \hfill
    \begin{subfigure}{0.33\linewidth}
        \centering
        \includegraphics[width=1\linewidth]{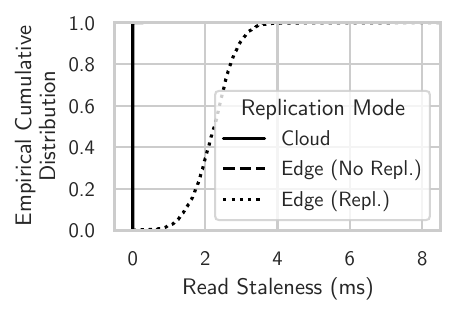}
        \caption{Data Staleness}
        \label{fig:replication:staleness}
    \end{subfigure}%
    \caption{Replication Results: Writing to a local data copy at the edge is faster than writing to the cloud (\cref{fig:replication:write}). Reading from a peer edge node is better that reading from the cloud in this scenario, yet reading from a locally replicated data copy is best (\cref{fig:replication:read}). The cost for this replication is data staleness (\cref{fig:replication:staleness}).}
    \label{fig:replication}
\end{figure*}

The read function results (\cref{fig:scaling:read}) illustrate two effects:
First, there is an overhead for each read that limits achievable throughput, especially for smaller data items where our requests do not saturate the ability of the system.
Second, the impact of the bandwidth constraints between edge and cloud are reflected in throughput achievable when deploying the data store in the cloud.
For data sizes >100kB, cloud store throughput hits the 12.5MB/s (100Mb/s) ceiling.

The throughput for our write-only workload (\cref{fig:scaling:write}) does not hit this limit, yet we can still observe that the more limited bandwidth and increased round-trip time negatively affect the performance of data updates when the data store for our function is located in the cloud compared to the edge.

\subsection{Replication}
\label{sec:evaluation:replication}

While the benefits of accessing data locally at the edge rather than at a remote cloud are obvious, a salient feature of \sysname{} is data replication across multiple edge nodes.

\subsubsection*{Setup}

To evaluate the impact of replication in \sysname{} on function performance, we extend our experiment setup of \cref{sec:evaluation:single} with a second edge node.
We emulate a 20ms network delay and 100Mb/s bandwidth limit between the two edge nodes.
We then deploy a function that can read and update a data item to both edge nodes and have the client update data on one node and read it from the other, measuring request-response latency and read staleness.
As shown in \cref{fig:replsetup}, we first have both function instances access a single cloud data store.
Second, we use a single data store only on the edge node where data is updated.
Read operations on the other edge node refer to this peer, similarly to the architecture of \emph{SyncMesh}~\cite{habenicht2022syncmesh}.
Third, we replicate data between the two nodes with \sysname{} to have both functions access data locally.
We consider a value as stale if it has been overwritten before the client reads it, with staleness measured as difference between current (read) time and timestamp of the operation that changed the value.
Note that we use the same client to read and write data in order to measure without clock drift.
Our client again sends ten requests per second for two minutes, with each experiment repeated three times.

\subsubsection*{Results}

The results in \cref{fig:replication} show an impact of data store location on write (\cref{fig:replication:write}) and read (\cref{fig:replication:read}) performance.
Writing to our cloud takes 50ms longer than updating data locally at the store integrated with \sysname{}.
Similarly, reading from a local data copy is 20ms and 50ms faster than reading from the peer edge and cloud nodes, respectively.
Interesting here is the price paid for this faster data access in data staleness (\cref{fig:replication:staleness}).
While data read and written to a single node with no replication does not lead to any staleness, we can see a median 2ms staleness with replication between edge nodes.
Note that read and write latency as well as read and writes not being at the exact same time partially hide replication delay, making the observable data staleness lower than the maximum 10ms one-way delay between edge nodes.
Also note that we here focus on staleness only as FReD already provides client-centric ordering guarantees.

\section{Multi-Function Application}
\label{sec:befaas}

\begin{figure}
    \centering
    \includegraphics[width=\linewidth]{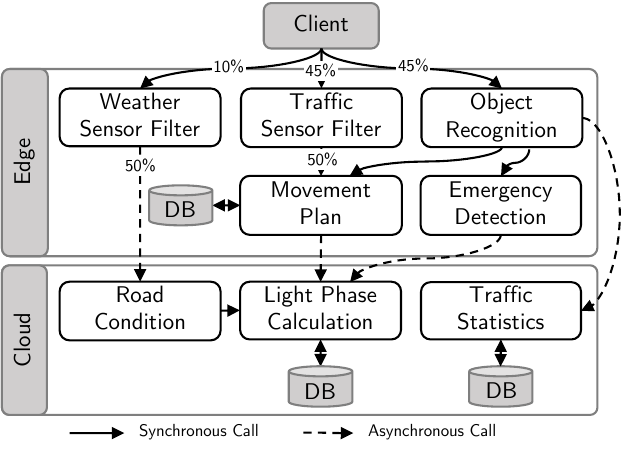}
    \caption{A multi-function smart city FaaS application adapted from the  BeFaaS benchmark suite~\cite{paper_grambow2021_befaas}.}
    \label{fig:befaas-setup}
\end{figure}

\begin{figure*}
    \centering
    \begin{subfigure}{0.33\linewidth}
        \centering
        \includegraphics[width=1\linewidth]{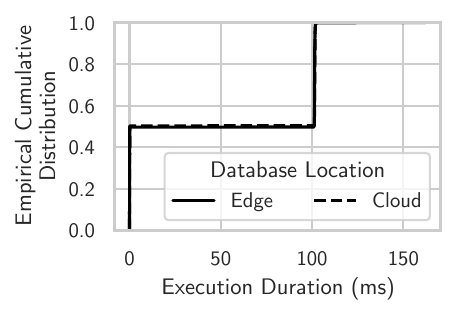}
        \caption{Weather Sensor Filter}
        \label{fig:befaas:weathersensorfilter}
    \end{subfigure}%
    \hfill
    \begin{subfigure}{0.33\linewidth}
        \centering
        \includegraphics[width=1\linewidth]{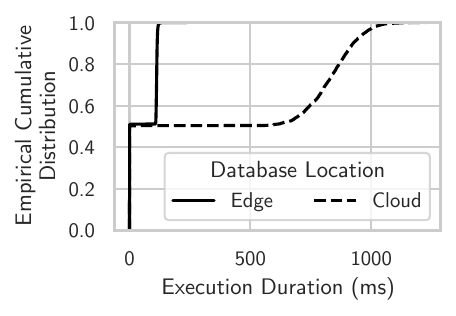}
        \caption{Traffic Sensor Filter}
        \label{fig:befaas:trafficsensorfilter}
    \end{subfigure}%
    \hfill
    \begin{subfigure}{0.33\linewidth}
        \centering
        \includegraphics[width=1\linewidth]{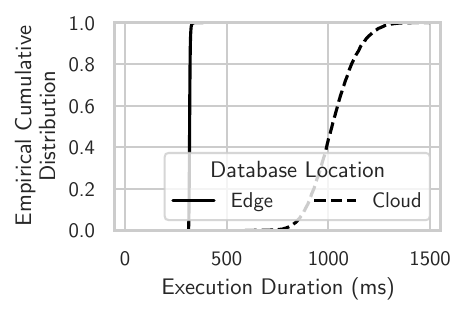}
        \caption{Object Recognition}
        \label{fig:befaas:objectrecognition}
    \end{subfigure}%
    \caption{The request-response latency measured in our edge-cloud FaaS application shows that local data access improves performance for functions that indirectly depend on data in their call chain.}
    \label{fig:befaas}
\end{figure*}

Finally, we show how a larger edge-cloud FaaS application composed of multiple functions can benefit from \sysname{}.

\subsubsection*{Setup}

We adapt an application from the \emph{BeFaaS}~\cite{paper_grambow2021_befaas} suite of application-centric FaaS benchmarks.
The smart city application shown in \cref{fig:befaas-setup} comprises eight FaaS functions distributed across the edge-cloud continuum that invoke each other synchronously or asynchronously.
Three functions persist data across function invocations in some database.

The client sends five request per second for a duration of ten minutes.
The client picks randomly between the \emph{traffic sensor filter} (45\%), \emph{object recognition} (45\%), and \emph{weather sensor filter} (10\%) functions.
The filter functions have a 50\% chance of calling subsequent functions, depending on input data.

We deploy this application across \sysname{} installations on edge and cloud hosts using the infrastructure configuration of \cref{sec:evaluation:single}.
We show the impact of replicating function state to the edge by comparing the deployments of the data store in cloud and edge.
We repeat this experiment three times.

\subsubsection*{Results}

We present request-response latency measurements from the perspective of the client in \cref{fig:befaas}.
As the weather sensor filter makes no synchronous requests to a function that depends on data, we see no impact of database location in \cref{fig:befaas:weathersensorfilter}.
As expected, the 50\% chance of an event being filtered by this function leads to a bimodal distribution of measured response latency.

The other two functions directly depend on the movement plan function, which accesses the key-value store multiple times during an invocation.
The request-response latency of the 50\% of requests that pass the traffic sensor filter (\cref{fig:befaas:trafficsensorfilter}) is significantly increased by the round-trip time between edge and data store.
Requests to the object recognition endpoint are similarly affected, as shown in \cref{fig:befaas:objectrecognition}.

\section{Related Work}
\label{sec:relwork}

FaaS has commanded significant research interest as an application paradigm in edge and cloud computing~\cite{raith2023serverless,jonas2019cloud,paper_pfandzelter2020_tinyfaas,palade2019evaluation,rausch234779,gackstatter2022pushing,raith2023serverless}.
Existing research has identified statelessness of FaaS functions as one of the key inhibitors to more broad adoption of this paradigm.
For example, Jonas et al.~\cite{jonas2019cloud} find that no existing cloud storage services are sufficient to provide a storage backend for FaaS as they lack low latency data access (block storage), cost efficiency (key-value databases), or persistence (in-memory storage).

Burckhardt et al.~\cite{burckhardt2021durable} introduce \emph{Durable Functions} (integrated with Microsoft Azure Functions), semantics for stateful FaaS computing based on the actor model.
With \emph{\textmu{}Actor}, Hetzel et al.~\cite{hetzel2021muactor} extend this actor model to the edge-cloud with content-based networking.
In contrast to our work, this programming model does not aim at reducing access latency and cost for FaaS applications but rather introduces synchronization and verifiable fault-tolerance on top of distributed, unreliable cloud FaaS platforms.

The need for native support of synchronization mechanisms in edge-cloud FaaS, e.g., for functions running longer than the typical 15-minute timeout of FaaS platforms has also been advocated for by Raith et al.~\cite{raith2023serverless}.
Karhula et al.~\cite{karhula2019checkpointing} introduce a checkpointing mechanism for edge FaaS that allows migrating long-running functions transparently to applications.
Such mechanisms, however, do not allow building applications that retain state across function invocations.

A key benefit of integrating storage with FaaS is overcoming the FaaS `data shipping' architecture:
The interaction with a central storage system, especially considering the network latency and bandwidth constraints between edge and cloud, are prohibitive to building efficient stateful edge applications on FaaS~\cite{aslanpour2021serverless}.
To this end, Rausch et al.~\cite{RAUSCH2021259} use application annotations to schedule edge containers close to their data sources.
Habenicht et al.~\cite{habenicht2022syncmesh} present \emph{SyncMesh}, a distributed FaaS platform for the edge where each node can keep local state in proximity to edge functions.
Instead of replicating data across edge nodes, data is only synchronized on-demand.
While useful for data processing applications, e.g., for the IoT, this introduces considerable latency in interactive applications where data synchronization between edge nodes lies on the hot path of function invocation.

Smith et al.~\cite{smith2022fado} implement function shipping in edge-cloud FaaS with \emph{FaDO}.
FaDO organizes data in buckets that can be replicated across multiple edge-cloud clusters.
Requests are sent to a central orchestrator that then routes them to a suitable function location based on data location.
This is a well-suited approach for data-intensive workloads, such as parallelized data processing, across cloud and on-premise `heavy' edge data centers.
With \sysname{}, in contrast, we target applications where clients directly access their local, lightweight edge FaaS instances to achieve lower execution latency than a central orchestrator could achieve.
With replication of both functions \emph{and} data at many edge locations, we thus ship neither functions nor data during execution but instead pay the price of replication in staleness.

\section{Conclusion \& Future Work}
\label{sec:conclusion}

In this paper, we have presented \sysname{}, an architecture for stateful and distributed FaaS for the edge-cloud continuum.
\sysname{} transparently integrates a FaaS platform with local data replication at each edge and cloud node, improving performance for edge functions without sacrificing the high level of abstraction of the FaaS paradigm.
In extensive evaluation with an open-source proof-of-concept prototype, we have shown the benefits of this approach compared to centralized data storage in the cloud and on-demand data loading from peer edge nodes.
We have also demonstrated how a larger edge-cloud FaaS application comprising multiple functions could benefit from this design.

In future work, we plan to further investigate the downsides of data replication, e.g., data staleness.
We also aim to provide recommendations to transparently manage these challenges in FaaS applications.

\begin{acks}
    Funded by the \grantsponsor{BMBF}{Bundesministerium f\"ur Bildung und Forschung (BMBF, German Federal Ministry of Education and Research)}{https://www.bmbf.de/bmbf/en} -- \grantnum{BMBF}{16KISK183}.
\end{acks}

\bibliographystyle{ACM-Reference-Format}
\bibliography{bibliography.bib}

\end{document}